\documentclass[12pt, a4paper]{article}
\usepackage[margin=1in]{geometry}
\usepackage[utf8]{inputenc}
\usepackage{latexsym}
\usepackage{changepage}  
\usepackage{hyperref}
\usepackage{csvsimple}
\usepackage{titlesec}
\titlelabel{\thetitle.\quad}

\makeatletter
\renewcommand{\@seccntformat}[1]{~\csname the#1\endcsname{}.}
\makeatother

\newcommand{\Section}{\section}

\makeatletter
\csvset{
  my hack/.style={
    late after line=\\\hline,
    late after last line=\csv@tablefoot\end{tabular}\csv@posttable},
}
\makeatother

\begin{document}

\begin{center}
    
    \Large\textbf{Performance evaluation of approximation algorithms for the minimum size 2-vertex strongly connected subgraph problem}   \\[18pt]
    
    \normalsize{Azzam Habib}\\
    
       \normalsize{\noindent Faculty of Informatics Engineering, Tishreen University, Syria}\\
    
    \rule{\textwidth}{0.6pt}
    \begin{abstract}
 Jaberi \cite{RaedJaberi} presented approximation algorithms for the problem of computing a minimum size 2-vertex strongly biconnected subgraph in directed graphs. We have implemented approximation algorithms presented in \cite{RaedJaberi} and we have tested the implementation on some graphs. The experimental results show that these algorithms work well in practice. 
    \end{abstract}
    \textbf{\normalsize {keywords}\\}
    \small{ Graph algorithms, Approximation algorithms, Connectivity}
    \rule{\textwidth}{0.4pt}
\end{center} 
\Section{Introduction}
If we ignore the directions of edges in a directed graph $G$, we obtain an undirected graph, called the underlying graph of $G$. A directed graph $G=(V,E)$ is strongly biconnected if for any two vertices $s,t \in V$, the directed graph $G$ contains a path from $s$ to $t$ and a path from $t$ to $s$ and the underlying graph of $G$ is biconnected. This concept was introduced by Wu and Grumbach \cite{ZhilinWuStephaneGrumbach}. A strongly biconnected graph $G=(V,E)$ is 2-vertex strongly biconnected if the removal of any vertex $v \in G$ leaves a strongly biconnected graph.

It is well known that the minimum $k$-vertex-connected spanning subgraph problem in directed graphs is NP-hard for $k\geq 1$ \cite{GareyDavidJohnson,JosephCheriyanRamakrishnaThurimella}. The edges set of each minimal $k$-vertex-connected directed graph contains at most $2kn$ edges \cite{JosephCheriyanRamakrishnaThurimella,JEdmonds,1985WMader}. Cheriyan and Thurimella \cite{JosephCheriyanRamakrishnaThurimella} gave  approximation algorithms with an approximation factor of $(1+1/k)$ for the minimum $k$-vertex-connected spanning subgraph problem. Georgiadis et al. \cite{LoukasGeorgiadisGiuseppeItalianoAikateriniKaranasiou,LoukasGeorgiadis} gave improved versions of these algorithms for the case $k=2$. The 2-vertex connected directed graphs have the property that they do not have any strong articulation point. Strong articulation points can be found using linear time algorithms presented by Italiano et al. \cite{GiuseppeItalianoLuigiLauraFedericoSantaroni}.
Jaberi \cite{RaedJaberi} presented approximation algorithms for the problem of computing a minimum size 
2-vertex strongly biconnected subgraph in directed graphs. We have implemented approximation algorithms presented in \cite{RaedJaberi} and we have tested the implementation on some graphs. The experimental results show that these approximation algorithms work well in practice.
\section{Approximation algorithms presented in \cite{RaedJaberi}}
In this section we review the approximation algorithms presented in \cite{RaedJaberi}.
Algorithm 1 shows the pseoducode of the first algorithm presented by Jaberi \cite{RaedJaberi} which is based on b-articulation points and minimal 2-vertex connected directed graphs.

\rule{17cm}{0.2mm}\\
\textbf{Algorithm 1} \cite{RaedJaberi}\\
\rule{17cm}{0.2mm}\\
\textbf{Input}: a 2-vertex strongly biconnected directed graph $G=(V,E)$\\
\textbf{Output}: a directed subgraph of $G$ which is $2$-vertex strongly biconnected\\
1.identify a minimal 2-vertex connected spanning subgraph $G^{+}=(V,E^{+})$\\
2. \textbf{if} $G^{+}$ has no b-articulation points \textbf{then}\\
3.  \space\space \space \space  output $G^{+}$\\
4. \textbf{else}\\
5. \space\space \space \space Let $K$ be set of the b-articulation points in $G^{+}$ \\
6. \space\space \space \space \textbf{for} each vertex $v \in K$ \textbf{do}  \\
7. \space\space \space \space \space\space \textbf{ while} $v$ is b-articulation point in $G^{+}$ \textbf{do} \\
8. \space\space \space \space \space\space   \space\space\space find and edge $(w.x)\in E\setminus E^{+}$ such that the strongly biconnected\\
9. \space\space \space \space \space\space  \space\space\space  component of $w$ does not contain the vertex $x$\\
10. \space\space \space \space \space\space  \space\space\space add the edge $(w.x)$ to $G^{+}$\\
11. \space\space \space \space output $G^{+}$\\
\rule{17cm}{0.2mm}\\\\
Jaberi \cite{RaedJaberi} proved that Algorithm 1 has an approximation factor of $(l/2+2)$, where $l$ is the number of the b-articulation points in the directed subgraph identified in line 1 of Algorithm 1 and Algorithm 1 runs in $O(n^{2}m)$ time.
   
Let $G =(V, E)$ be a 2-vertex strongly biconnected graph, we say that $G$ is minimal if $G\setminus \lbrace e\rbrace$ is not 2-vertex strongly connected for any edge $e\in E$. Jaberi \cite{RaedJaberi} also proved that each minimal 2-vertex strongly biconnected directed graph has at most $7n$ edges. His proof is based on the results of Mader and  \cite{1971WMader,1972WMader,1985WMader} and Edmonds \cite{JEdmonds}. This imply a 7/2 approximation algorithm for the problem of computing a minimum size 2-vertex strongly biconnected subgraph in directed graphs.   

In order to implement this algorithm, we used the simplest approach, just by removing edges one by one and check if the remaining graph is strongly biconnected. Algorithm 2 shows the pseudocode of this simple approach which runs in $O(nm(n + m))$ time.

\rule{17cm}{0.2mm}\\
\textbf{Algorithm 2}\\
\rule{17cm}{0.2mm}\\
\textbf{Input}: A 2-vertex strongly biconnected directed graph $G=(V,E)$\\
\textbf{Output}: A minimal $2$-vertex strongly biconnected spanning subgraph of $G$\\
1. \textbf{for} each edge $e\in E$ \textbf{do}\\
2.\space \space \space\space \textbf{if} $G\setminus \lbrace e\rbrace$ is 2-vertex strongly connected \textbf{then} \\
3.\space \space \space\space \space\space remove $e$ from $G$\\
4.Output $G$\\
\rule{17cm}{0.2mm}\\\\

 Cheriyan and Thurimella \cite{JosephCheriyanRamakrishnaThurimella} gave  approximation algorithms with an approximation factor of $(1+1/k)$ for the minimum $k$-vertex-connected spanning subgraph problem. The output of this algorithm is not necessarily $2$-vertex strongly biconnected when $k=2$. Jaberi \cite{RaedJaberi} described a modified version of this algorithm which is able to obtain a $2$-vertex-strongly biconnected subgraph. We have not implemented this version but we have implemented a modified version which finds a matching in the first phase using a greedy algorithm instead of finding a $\geq 1$ matching. We refer to the simple version that finds a matching using a greedy algorithm as Algorithm 3.
 \section{The results of our experiments}   
Here we report the results of the experiments that we conducted. 
 
\begin{tabular}{||c||c||c||c||c||c||c|}
   \hline 
   Input & Algorithm2 & Algorithm2 & Algorithm3 & Algorithm3& Algorithm1 & Algorithm1 \\ 
   \hline 
   ( V , E ) & Time & Edges & Time & Edges & Time & Edges \\ 
   \hline 
   ( 10 , 54 ) & 1 s & 21 & 1 s & 22 & 1s & 21 \\ 
   \hline 
   ( 50 , 387 ) & 8 s & 113 & 8s & 111 & 1s & 113 \\ 
   \hline 
   ( 100 , 1000 ) & 18 s & 226 & 16s & 226 & 2s & 227 \\ 
   \hline 
   ( 200 , 2000 ) & 2 m & 450 & 1 m 38 s & 457 & 18s & 451 \\ 
   \hline 
   ( 300 , 2724 ) & 14 m 26 s & 679 & 8 m 4 s & 676 & 57 s & 679 \\ 
   \hline 
   ( 500 , 4914 ) & 36 m 31 s & 1145 & 34 m 7 s & 1134 & 5 m 41 s & 1145 \\ 
   \hline 
   \end{tabular}

We implemented all our algorithms in Java without the use of any external graph library. The experiments were conducted on a 64-bit Intel(R)/Windows machine running Windows 21H1, with a 1.99 GHz Intel i7- 8550U, 12 GB of RAM, 8 MB of L3 cache, and each core has a 256 KB private L2 cache. All experiments were executed on a single core without using any parallelization. We report CPU times measured with the nanoTime function.
For our experimental study, we used a collection of random generated graphs, we constructed 2-vertex strongly biconnected graphs as follows. We generated random graphs with n node and 3n random edges, then we keep adding edges one by one until the graph become 2-vertex strongly biconnected. For these graphs, notice that the output of the algorithms presented in \cite{RaedJaberi} has less than $3n$ edges.

\end{document}